\documentclass[runningheads]{llncs}

\usepackage{semantic}
\usepackage{fontspec}
\usepackage{newunicodechar}
\usepackage[nounderscore]{syntax}

\usepackage{caption}
\usepackage{subcaption}
\usepackage{proof}
\usepackage{xcolor}
\usepackage{amsmath}
\usepackage{amsfonts}
\usepackage[ruled]{algorithm2e}
\usepackage{hyperref}
\usepackage{url}
\usepackage{placeins}

\newcommand{\term}[2][]	{\ifthenelse{\equal{#1}{}}{\index{#2}}{\index{#1}}\emph{#2}}
\newcommand{\heading}[1] {\noindent {\bf #1}}

\newcommand{\code}[1]{\texttt{#1}}

\newcommand{\BD}{\begin{definition}}
\newcommand{\ED}{\end{definition}}

\newcommand{\intersect}{\cap}

\newcommand{\idx}[2][]	{\ifthenelse{\equal{#1}{}}{\index{#2}#2}{\index{#1}#2}}

\newcommand{\eaci}[1]{\edef\tmp{\noexpand\index{\ac{#1}}} \emph{\tmp}}

\newcommand{\defvoc}[2]{\expandafter\def\csname #1\endcsname{#2} {\newacro{#1}{#2}}}

\newcommand{\vf}[1]{\def\tmp{\index{\csname #1\endcsname@{\acf{#1}}}}\tmp\emph{\acf{#1}}}
\newcommand{\vs}[1]{\def\tmp{\index{\csname #1\endcsname@{\acf{#1}}}}\tmp\acs{#1}}
\newcommand{\vl}[1]{\def\tmp{\index{\csname #1\endcsname@{\acl{#1}}}}\tmp\acl{#1}}

\SetKw{Throw}{throw}

\usepackage{listings}

\usepackage{color}
\definecolor{keywordcolor}{rgb}{0.7, 0.1, 0.1}   
\definecolor{tacticcolor}{rgb}{0.0, 0.1, 0.6}    
\definecolor{commentcolor}{rgb}{0.4, 0.4, 0.4}   
\definecolor{symbolcolor}{rgb}{0.0, 0.1, 0.6}    
\definecolor{sortcolor}{rgb}{0.1, 0.5, 0.1}      
\definecolor{attributecolor}{rgb}{0.7, 0.1, 0.1} 

\newunicodechar{→}{\ensuremath{\to}}\newunicodechar{→}{\ensuremath{\to}}
\newunicodechar{ℝ}{\ensuremath{\mathbb{R}}}
\newunicodechar{𝕜}{\ensuremath{\mathbb{k}}}
\newunicodechar{𝓝}{\ensuremath{\mathcal{N}}}
\newunicodechar{∀}{\ensuremath{\forall}}
\newunicodechar{∃}{\ensuremath{\exists}}
\newunicodechar{ℤ}{\ensuremath{\mathbb{Z}}}
\newunicodechar{λ}{\ensuremath{\lambda}}
\newunicodechar{α}{\ensuremath{\alpha}}
\newunicodechar{β}{\ensuremath{\beta}}
\newunicodechar{τ}{\ensuremath{\tau}}
\newunicodechar{∘}{\ensuremath{\circ}}
\newunicodechar{≠}{\ensuremath{\neq}}
\newunicodechar{↦}{\ensuremath{\mapsto}}
\newunicodechar{∧}{\ensuremath{\wedge}}
\newunicodechar{₁}{\ensuremath{_1}}
\newunicodechar{₂}{\ensuremath{_2}}
\newunicodechar{₁}{\ensuremath{_1}}
\newunicodechar{₃}{\ensuremath{_3}}
\newunicodechar{¹}{\ensuremath{^1}}
\newunicodechar{⁻}{\ensuremath{^-}}

\title{Modular Composition of Inductive Types Using Lean Meta-programming}
\author{Ramy Shahin\orcidID{0000-0001-8724-3934}}
\institute{Qualgebra\\
\email{ramy@qualgebra.com}} 

\authorrunning{R. Shahin}

\titlerunning{Modular Composition of Inductive Types Using Lean Meta-programming}

\begin{document}

\maketitle

\begin{abstract}
    Inductive types are ubiquitous building blocks in many programming and theorem proving languages. 
    An inductive type is a closed set of constructors from which values of the type can be created. 
    That set cannot be extended though once a type is defined. 
    This limits extensibility, reuse, and modular separation of concerns when defining types and functions operating over their values. 
    This limitation is manifested in the expression problem, where extending an expression language with new syntactic constructors without having to modify or re-compile existing ones is a challenge in almost all programming languages.
    
    This paper presents inductive type and function implementation composition algorithms based on meta-programming.
    In addition, a set of syntactic extensions to the Lean proof assistant implementing those algorithms are presented. 
    This framework allows for modular reuse, composition, and extension of a subset of Lean type and function definitions. 
    In addition, semantic subtyping relations between component and composite types are discussed both at the type theoretic and implementation levels.

    The framework is demonstrated on a case study, involving the composition of syntactic and semantic artifacts of three sublanguages into one language. 
    The case study highlights both the features and limitations of the composition framework.

\end{abstract}

\section{Introduction}
\label{sec:intro}
\vspace{-0.15in}

Interactive theorem provers are slowly gaining academic and industrial adoption as quality assurance and automated reasoning tools for software systems.
Ensuring the correctness of software, especially for safety-critical and mission-critical systems, has been a major challenge for decades.
Among other approaches, such as model checking, static analysis, and automated testing, interactive theorem proving has several success stories in the last 20 years.
Examples include the certified CompCert C compiler~\cite{Leroy:2009} and the certified seL4 microkernel~\cite{Klein:2009}.

However, one of the reasons this adoption has been slow is the amount of effort needed to define and prove correctness theorems pertaining to software systems and components.
Parallel to software engineering, a discipline of proof engineering has been evolving, trying to adapt software engineering best practices such as decomposition, reuse, modularity, and encapsulation to theorem proving languages.
Although the analogy between software and proof artifacts still has some intricate gaps, software engineering techniques (at least at a high level) can still tackle the inherent complexity in developing certified programs. 

Data types and logical predicates are typically defined inductively in theorem provers.
Inductive types are at the core of many programming and theorem proving languages because they come with automatically generated induction principles.
Proofs of theorems defined on those types and predicates can use those induction principles directly, or use induction tactics which automate some steps in the process.

An inductive type is a closed set of constructors from which values of the type can be created. 
That set of constructors cannot be extended though once a type is defined.
In the absence of modular composition constructs, extending a theory based on an inductive type typically involves code redundancy.
This limits extensibility, reuse, and modular separation of concerns when defining types and functions operating over their values.
One manifestation of this problem is the \term{expression problem}~\cite{Wadler:1998}, where extending an expression language with new syntactic constructors while reusing existing ones without having to modify or re-compile them is a challenge in almost all programming languages.
Attempts at tackling this problem both in programming and theorem proving languages include~\cite{Boite:2004,Delaware:2011,Swierstra:2008,Delaware:2013,Forster:2020,Jin:2023}.

\heading{Motivating Example. }
When formalizing the syntax and semantics of a logic such as Kleene's K3 logic~\cite{Hazen:2019} in a theorem prover, the entire logic is typically formalized as one monolithic model.
Fig.~\ref{fig:monolithicK3} shows an example of such a model of syntactic terms of K3, and a function to count the number of Abstract Syntax Tree (AST) nodes of those terms.
However, if a model of propositional logic already exists, in principle it should be reused and extended only with the indeterminate truth value \code{I}, with respective extensions to the different functions and predicates operating over terms.

Furthermore, the negation, conjunction, disjunction, and implication logical connectives form by themselves a component reusable in different logics.
If the connectives are packaged as a separate component (e.g., Fig.~\ref{fig:ops}), they can be composed with Boolean truth values (Fig.~\ref{fig:bool}) to build a model of propositional logic.
If a component modeling the K3 indeterminate value (Fig.~\ref{fig:indet}) is added to both components, the resulting composite is equivalent to the monolithic K3 model.
Similarly, if components modeling other third values in three-valued logics are composed with the components of Boolean value and connectives, we end up with models of other three-valued logics.

Including only a single coherent logical aspect in a component makes it both simpler to develop and easier to reuse. 
Inductive data types are the building blocks of programs and proofs in Lean, Rocq, and similar languages and proof assistants.
Inductive types are closed though, i.e., not extensible.

Closed inductive types are central to the Calculus of Inductive Constructions (CIC)~\cite{Pfenning:1990}, which is the core calculus of Lean and Rocq.
To avoid compromising the soundness of the logic (and the languages and tools built on top of it) by adding syntactic constructs for modularity and compositionality to the core language, meta-programming is a safer alternative as it does not change the core language or inflate the Trusted Computing Base (TCB).

This paper presents a set of syntactic extensions to the Lean programming language that allow the modular compositions of inductive types, together with functions and proofs operating on their values.
Lean 4~\cite{deMoura:2021} includes meta-programming constructs that allow developers to extend the syntax of the language, and provide user-defined elaborators of the extended syntactic constructs.
Those meta-programming facilities are utilized to allow the definition of inductive types that are built from the constructors of component types, and pattern matching cases in function definitions.
Our composition approach supports both recursive inductive types and recursive functions. 
In addition, the composition framework automatically generates coercion operators that allow passing values of component types to functions expecting values of the composite type.

The design decisions of this framework were guided by three goals: (1) to allow the composition of plain Lean type and function definitions, without requiring extensional encodings of those definitions in some frameworks (similar to the types a-la-carte line of work~\cite{Swierstra:2008,Delaware:2013}), (2) to prioritize usability over generality, focusing on the fragment of Lean covering polymorphic dependent types, instead of trying to cover the entire language, and (3) to reuse parts of existing definitions that are free of feature interactions, while allowing users to handle feature interaction cases.

\heading{Contributions and Outline. }
This paper makes the following contributions:
(1) The design of meta-programming algorithms for composing inductive data types, functions pattern matching on their values, and generating implicit coercion operators from component to composite types.
(2) An open-source prototype implementation of those algorithms on top of the Lean meta-programming framework. This includes syntactic operators for composing inductive types and functions\footnote{\url{https://github.com/qualgebra/LeanToolkit/tree/SEFM2026}}.
(3) A case study on how our Lean implementation can be used to compose definitions of syntactic languages, functions, predicates, theorems, and proofs over their terms. This case study helped us identify some limitations of the framework, which we also discuss.

Following this introduction, a brief background on Lean, meta-programming, and proof engineering is presented (Sec.~\ref{sec:background}).
The meta-programming algorithms for composing inductive types (Sec.~\ref{sec:typeComp}) and functions (Sec.~\ref{sec:funComp}) are then covered.
A case study is then presented in Sec.~\ref{sec:casestudy}, together with a discussion of the limitations of the framework identified during its implementation.
We then briefly survey some related work (Sec.~\ref{sec:related}), and finally conclude.

\begin{figure}[t]
    \vspace{-0.15in}
    \begin{subfigure}{0.49\textwidth}
        \begin{lstlisting}
inductive Term
| T | F | I
| Neg (t: Term)
| And (t1 t2: Term)
| Or (t1 t2: Term)
| Imp (t1 t2: Term)

def countNodes: Term→ Nat
| .T => 1
| .F => 1
| .I => 1
| .Neg t => 1 + countNodes t
| .And t1 t2 => 1 + countNodes t1 + countNodes t2
| .Or t1 t2 => 1 + countNodes t1 + countNodes t2
| .Imp t1 t2 => 1 + countNodes t1 + countNodes t2
        \end{lstlisting}
        \vspace{-0.2in}
        \caption{Monolithic modeling of K3.}
        \label{fig:monolithicK3}
    \end{subfigure}
    \hfill
    \begin{subfigure}{0.49\textwidth}
        \begin{lstlisting}
namespace Ops

inductive Term
| Neg (t: Term)
| And (t1 t2: Term)
| Or (t1 t2: Term)
| Imp (t1 t2: Term)

def countNodes: Term → Nat
| .Neg t => 1 + countNodes t
| .And t1 t2 => 1 + countNodes t1 + countNodes t2
| .Or t1 t2 => 1 + countNodes t1 + countNodes t2
| .Imp t1 t2 => 1 + countNodes t1 + countNodes t2

end Ops
        \end{lstlisting}
        \vspace{-0.2in}
        \caption{Operators component.}
        \label{fig:ops}
    \end{subfigure}
    \hfill 
    \begin{subfigure}{0.49\textwidth}
        \begin{lstlisting}
namespace Boolean

inductive Term
| T
| F

def countNodes: Term→ Nat
| .T => 1
| .F => 1

end Boolean
        \end{lstlisting}
        \vspace{-0.2in}
        \caption{Boolean component.}
        \label{fig:bool}
    \end{subfigure}
    \hfill
    \begin{subfigure}{0.49\textwidth}
        \begin{lstlisting}
namespace Indet

inductive Term
| I

def countNodes: Term → Nat
| .I => 1

end Indet
        \end{lstlisting}
        \vspace{-0.2in}
        \caption{Indeterminate component.}
        \label{fig:indet}
    \end{subfigure}
    \vspace{-0.15in}
    \caption{A monolithic model of K3, vs. three separate modules that can be composed into K3 among other modular configurations.} 
    \label{fig:example}
    \vspace{-0.25in}
\end{figure}

\vspace{-0.15in}
\section{Background}
\label{sec:background}
\vspace{-0.15in}

This section introduces the concepts and notations used throughout the paper.

\heading{Inductive Data Types.}
An inductive data type \code{T} has a set of constructors ${\code{C}_1, ... , \code{C}_n}$.
A value of type \code{T} can only be created using one of those constructors. 
Each constructor can take zero or more typed arguments.
An inductive type \code{T} is recursive if at least one of its constructors has an argument of type \code{T}. 


\heading{Metaprogramming in Lean.}
Lean is an interactive theorem prover based on the Calculus of Inductive Constructions (CIC)~\cite{Pfenning:1990}.
The name Lean refers to both the theorem prover and the language in which definitions, theorems, and proofs are written.
Inductive data types, including inductive predicates, are among the main building blocks of the CIC and Lean.

Lean 4~\cite{deMoura:2021} includes an extensive meta-programming framework written in Lean itself.
This framework allows the development of proof tactics, syntax extensions, environment extensions and Domain Specific Languages (DSLs) in Lean.
Syntax extensions come in the form of user defined syntactic categories with grammar rules. 
Lean automatically generates parsers based on those rules, registers those parsers, and automatically invokes them when code patterns following the rules are encountered during parsing.
Parsing turns sequences of tokens into Lean \emph{terms}. 
When developing a syntax extension, a user-defined \emph{elaborator} processes those terms into Lean expressions.
Those expressions are then compiled by the Lean compiler into the minimalistic kernel language, which is mostly CIC constructs.
During compilation, the Lean type checker validates the logical soundness of proofs written in the source language.

\heading{Separation of Concerns and Proof Engineering.}
Theorem provers have been slowly integrated in the Software Development Life Cycles (SDLCs) of some safety-critical and mission-critical software projects over the years.
Because of the complexity of such systems, and software systems in general, decomposition of a system into smaller, manageable modules has been a foundational software engineering principle~\cite{Parnas:1972}.
Modules are expected to be cohesive, self-contained, and to hide details that are irrelevant to other modules. 
Another important principle of software engineering is separation of concerns~\cite{Dijkstra:1982}, where different aspects of a software system should not belong to the same module.

Several programming paradigms have been proposed to facilitate the separation of concerns at the code level, such as Subject-oriented Programming~\cite{Harrison:1993} and Aspect-oriented Programming~\cite{Kiczales:1997}.
The goal is to allow developers to design and implement self-contained modules separately, and then compose them in effective ways and reusing them in different contexts and projects.

In the CIC, theorems are types, and proofs are terms. 
Theorem prover languages such as Gallina and Lean already come with powerful type systems including features such as type polymorphism, typeclases, and operator overloading.
However, copying-and-pasting code is probably still the most commonly used reuse technique, especially for type definitions because inductive types are closed and not extensible.

A software feature is typically a unit of functionality. 
A system is typically composed of multiple features.
Each feature has a set of properties expected by users and intentionally designed by engineers.
However, in many cases one feature might affect the behavior of another in a system, possibly violating its intended properties.
This is usually referred to as a \term{feature interaction}~\cite{Batory:2011}.


\vspace{-0.15in}
\section{Composing Types}
\label{sec:typeComp}
\vspace{-0.15in}



Composing a set of inductive types results in a composite type that includes each of the constructors in its components.
For example, to compose \code{Boolean.Term}, \code{Indet.Term}, and \code{Ops.Term} (Fig.~\ref{fig:example}) into a new type \code{K3Term}, the composition framework supports the following Lean extended syntax:
\begin{lstlisting}[firstline=143,lastline=143]
inductive K3Term := Boolean.Term |+ Indet.Term |+ Ops.Term
\end{lstlisting}

Running this command generates and elaborates the following command:
\begin{lstlisting}
inductive K3Term
| T: K3Term 
| F: K3Term 
| I: K3Term
| Neg: K3Term → K3Term
| And: K3Term → K3Term → K3Term
| Or: K3Term → K3Term → K3Term
| Imp: K3Term → K3Term → K3Term 
\end{lstlisting}

Note that recursive constructors in the composite type now have arguments of the type \code{K3Term} instead of \code{Ops.Term}.
The semantics of the type composition operator \code{|+} are captured by the $\texttt{cns|+}$ inference rule:
\[
    \inference 
        {\code{S:=T$_0$|+...|+T$_m$} \quad \code{C}:\tau_0\to\dots\to\tau_n\to\code{T$_i$}\in ctors(\code{T$_i$}), 0\le i\le m} 
        {\code{C}: \emph{adjust}(\tau_0) \to ... \to \emph{adjust}(\tau_n) \to \code{S} \in \emph{ctors} \code {(S)}}
        [cns|+]
\]
\[
    \emph{where}~\emph{adjust}(\tau) = \code{if $\tau \in$ \{T$_0$, ..., T$_m$\} then S else $\tau$}
\]
When composing component types \code{T$_0$, ..., T$_m$} into composite type \code{S} (written as \code{S := T$_0$ |+ ... |+ T$_m$}), each constructor belonging to types \code{T$_0$, ..., T$_m$} also belongs to \code{S}, modulo type adjustment.
Given constructor \code{C $\in$ T$_i$} of type $\tau_0 \to \dots \to \tau_n \to \code{T}_i$, any recursive references to the types \code{T$_0$, ..., T$_m$} are replaced with references to the new type \code{S} instead.
Another kind of adjustment that deals with references to component types and functions that have been previously composed is discussed later in this section.

A precondition for type composition is that all the component types have the same higher order type. Composition fails with an error message if this precondition is not met. 
For example, \code{U: Type} and \code {V: Type $\to$ Type} are not composable.
Another precondition is that the sets of constructor names within the component types are disjoint, i.e., constructor names across the syntactic scopes of component types are unique.

Alg.~\ref{alg:sum} outlines the logic of composing two types $t_1$ and $t_2$ into a new inductive type with name $n$. 
Because type composition is associative, this algorithm can be used successively to compose sequences of types of arbitrary length.
An additional optional parameter of the algorithm is a set of extra constructors $cs$ that can be added to the ones in $t_1$ and $t_2$.

Type signatures of $t_1$ and $t_2$ are first checked for equivalence (lines 1-5).
The sets of constructors of types $t_1$ and $t_2$ are then retrieved (lines 6-7).
The second precondition of constructor name uniqueness among $t_1$, $t_2$, and the new constructors is then checked (lines 8-10).
If the checks of any of the two preconditions fail, an exception is thrown.

The name of the new type is then registered as a synonym for both $t_1$ and $t_2$ (lines 11-12).
The \code{ajustType} routine (Alg.~\ref{alg:adjust}) depends on those synonym definitions.
We then iterate through all the constructors of $t_1$ and $t_2$ and adjust their types using \code{adjustType}, inserting the adjusted constructors into set $c_m$ (lines 13-17).
A new inductive type is then created, with name $n$, signature $s_1$ (which is the signature of both types $t_1$ and $t_2$), and the set of constructors $c_m \cup cs$ (lines 18-19), and finally that type is returned.

\LinesNumbered

\begin{algorithm}[t]
\caption{Sum of two inductive types.} 
\label{alg:sum}
\KwIn{$t_1$, $t_2$: Type, $cs$: Constructors, $n$: Name}
\KwOut{Type}

$s_1 \gets \textrm{signature}(t_1)$\;
$s_2 \gets \textrm{signature}(t_2)$\;
\If{$s_1 \neq s_2$}
    {\Throw{"incompatible signatures"}}

$c_1 \gets \textrm{constructors}(t_1)$\;
$c_2 \gets \textrm{constructors}(t_2)$\;
\If{$\textrm{names}(c_1) \intersect \textrm{names}(c_2) \neq \emptyset \vee \textrm{names}(cs) \intersect (\textrm{names}(c_1) \cup \textrm{names}(c_2)) \neq \emptyset$}
    {\Throw{"conflicting constructor names"}}

$\textrm{registerSynonym}(n, name(t_1))$\;
$\textrm{registerSynonym}(n, name(t_2))$\;

$c_n \gets c_1 \cup c_2$\;
$c_m \gets \emptyset$\;

\For{$c \in c_n$}{
    $\textrm{insert}(c_m, \textrm{updateType}(c, \textrm{adjustType}(type(c))))$\; 
}

$c_m \gets c_m \cup cs$\;

$t \gets \textrm{createInductiveType}(n, s_1, c_m)$\;

\Return $t$
\end{algorithm}

The types of composed constructors need to be adjusted in two ways: First, if they contain any reference to the types being composed, those references have to be changed to the name of the new composite type $n$.
Second, if they refer to any types or functions that have been previously composed, those references also need to be changed to the names of the respective composite type/function names.
This is the reason why we need to register composite type names as synonyms for their constituents.
Here we make the simplifying assumption that a type will be composed at most once within a given module.

Alg.~\ref{alg:adjust} outlines the logic of the \code{adjust} routine. It takes a type $t$ as input, and returns a type.
It uses a subroutine \code{renameType} (lines 9-14), which takes a type name, checks if it has a synonym, and returns that synonym if it exists, otherwise returns its original input.
The \code{adjust} routine starts by pattern matching on the structure of its input $t$.
If it is the name of a type, that name is passed to \code{renameType}, and the result is returned directly.
On the other hand if it is an arrow type (or a $\forall$ dependent type), the head of that arrow chain is passed to \code{renameType}, and \code{adjust} is recursively called on the rest of the chain, which is a type itself.

\SetKwFor{Case}{case}{}{end case}
\SetKwComment{Comment}{/* }{ */}
\begin{algorithm}[t]
    \caption{Adjust the type of a constructor.} 
    \label{alg:adjust}
    \KwIn{$t$: Type}
    \KwOut{Type}
    \Switch{t} {        
            \Case{u}{
               \Return renameType(u) 
            }
            \Case {$u \to v$}{
                Return (renameType(u) $\to$ adjust(v))
            }
    }
    \SetKwProg{myproc}{Procedure}{}{}
    \myproc{renameType(t: Type)}{
    \eIf{hasSynonym(t)}
        {\Return synonym(t)}
        {\Return t}
    }
\end{algorithm}

\heading{Semantic Subtyping Relation.}
In the Calculus of Inductive Constructions (CIC), values created with different constructors are different.
However, \emph{semantically} \code{Boolean.Term.T} and \code{K3Term.T} denote the same thing, and using them interchangeably should be allowed whenever possible.
If \code{T} is one of the component types and \code{S} is the composite type, then \code{T <: S}, i.e., \code{T} is a semantic subtype of \code{S}.
\[
    \inference 
        {\code{S := T$_0$ |+ ... |+ T$_m$}}
        {\code{T$_0$ <: S}~\wedge~\dots~\wedge~\code{T$_m$ <: S}}
        [subtype|+]
\]
This semantic subtyping relation allows for passing values of component types to functions expecting values of the composite type (either composite functions or hand-written ones).

\heading{Coercion Operators.}
Encoding the semantic subtyping relation in Lean is accomplished by implementing coercion operators between component types and composite types.
Lean supports implicit coercion from subtypes to supertypes, and allows programs to define their own coercion operators by instantiating the \code{Coe} typeclass.
For each component type \code{T} in a composite type {S}, the composition framework automatically generates a typeclass instance of this form:
\begin{lstlisting}
instance: Coe T S where
    coe := fun (x: T) =>
                match x with
                | T.cns1 ... => S.cns1 ... 
                ... 
                | T.cnsn ... => S.cnsn ...
\end{lstlisting}

The body of the coercion function pattern matches over constructors of type \code{T}, and maps each of them to the corresponding constructor of type \code{S}. 
This is a straightforward translation because both constructors have the same signature (modulo type adjustment, where the coercion function is applied resursively), and only differ in their type name prefix.

For instance, the coercion operator from \code{Boolean.Term} to \code{K3Term} is automatically generated as
\begin{lstlisting}[xleftmargin=0pt,firstline=163,lastline=167]
def coe.Boolean.Term.K3Term : Boolean.Term → K3Term :=
fun x =>
  match x with
  | Boolean.Term.T => K3Term.T
  | Boolean.Term.F => K3Term.F
\end{lstlisting}

This operator is registered with Lean to get automatically invoked by instantiating the \code{Coe} typeclass as follows:
\begin{lstlisting}[xleftmargin=0pt,firstline=174,lastline=175]
def SubType.Boolean.Term.K3Term : SubType Boolean.Term K3Term :=
{ coe := coe.Boolean.Term.K3Term }
\end{lstlisting}

The \code{SubType} typeclass inherits from \code{Coe}, and is used exclusively for coercion instances between component and composite types.
In addition to providing a coercion operator, it also maps a component type to a composite type in the Lean environment.
This is used whenever a reference to a component type needs to be adjusted into its corresponding composite type.

\heading{Dependent Coercion Operators.}
The Lean \code{Coe} typeclass is used for \term{casting up}, from a value of a subtype to a value of a corresponding supertype.
Lean also supports coercion in the opposite direction, from constructor-specific values of a supertype down to a subtype.
The Lean \code{CoeDep} typeclass is used for this purpose.
The constructor used to create the value being coerced has to be explicitly included in the syntactic value, hence the \emph{dependent} nature of this coercion scheme.
For example, the following instance is automatically generated as a part of composing the \code{K3Term} type:
\begin{lstlisting}[xleftmargin=0pt,firstline=182,lastline=183]
def CoeDep.K3Term.T.Boolean.Term : CoeDep K3Term K3Term.T Boolean.Term :=
{ coe := Boolean.Term.T }
\end{lstlisting}

The coercion operator in this instance is automatically invoked whenever \code{K3Term.T} is used in a syntactic context where a value of type \code{Boolean.Term} is expected.
Similar instances are generated for all non-recursive constructors of \code{K3Term}.
Dependent coercion does not typecheck for recursive constructors. 
For example, \code{K3Term.And: K3Term → K3Term → K3Term}, so given arbitrary conjuncts \code{a} and \code{b}, \code{K3Term.And a b} cannot be coerced to \code{Ops.And a b} because the constructors used to create \code{a} and \code{b} are not know to the typechecker.
This limitation of dependent coercions implies that their invocations are less frequent than independent, up-cast ones.
However, as demonstrated in Sec.~\ref{sec:funComp}, they are useful when composing functions over component types.

\vspace{-0.15in}
\section{Composing Functions}
\label{sec:funComp}
\vspace{-0.15in}

To be able to effectively reuse component types, in addition to composing the types, we also need to compose functions operating over their values.
For example, the \code{countNodes} function in Fig.~\ref{fig:example} is implemented in each of the \code{Boolean}, \code{Indet}, and \code{Ops} namespaces.
Each of those functions pattern matches over the constructors of its respective \code{Term} type.
When composing the \code{Term} types from the three namespaces, we would ideally want to compose the three \code{countNodes} function implementations as well.

A function operating on an inductive type typically pattern matches on the structure of the different constructors of its input type.
When composing functions, we are effectively calculating the set union of disjoint sets of pattern matching alternatives coming from the component functions.
Function composition has two preconditions: 
First, the component functions should have the same signature, modulo the type(s) of the parameter(s) the function is pattern matching on.
The second precondition deals with those parameters the function is pattern matching on. The types of those parameters should themselves be included in their respective composite types.
\[
    \inference 
        {\code{f$_0$: T$_0 \to$ S} \quad \dots \quad \code{f$_n$: T$_n \to$ S} \quad \code{T := T$_0$ |+ ... |+ T$_n$}}
        {\code{(f$_0$ |+ \dots~|+ f$_n$): T $\to$ S}}
        [function|+]
\]
When composing pattern matching alternatives from different functions, the \code{adjust} routine (Alg.~\ref{alg:adjust}) is used to rename all references to component types and functions to their registered synonyms.
This implies that the order in which types and functions are composed within a code module should respect the interdependencies between those artifacts, i.e., if a composite function \code{f} depends on a composite type \code{T}, \code{T} should be composed first.

Note that a composite type \code{T} is a superset of \code{T$_0$ $\cup~\dots~\cup$ T$_n$} in the general case.
In other words, there might be values belonging to \code{T} but not to any of the \code{T$_i$} types.
For example, the value term \code{(K3Term.And K3Term.T K3Term.I)} belongs to \code{K3Term}, but is not a value of \code{Boolean.Term}, \code{Indet.Term}, or \code{Ops.Term}.
This is particularly true when composing types with recursive constructors.
When composing functions over those types, there will always be pattern matching cases not covered by any of the component functions.

For instance, when composing the three \code{countNodes} functions in Fig.~\ref{fig:example}, the pattern matching cases for the non-recursive \code{K3Term} constructors (\code{K3Term.T}, \code{K3Term.F}, and \code{K3Term.I}) are automatically generated. The pattern matching alternative for each simply calls the corresponding function, implicitly leveraging the dependent coercion operator for each of those constructors.
The \code{K3Term.T} pattern matching case for example is \code{Boolean.countNodes K3Term.T}.
Since the argument of \code{Boolean.countNodes} is of type \code{Boolean.Term}, while \code{K3Term.T} is of type \code{K3Term}, this plain term does not typecheck.
Because a dependent coercion instance is defined for this particular constructor, the dependent coercion operator is automatically invoked, replacing \code{K3Term.T} with \code{Boolean.Term.T}.

A value created using a recursive constructor on the other hand potentially cross-cuts the different component types, so a dependent coercion operator for such a value would not be accepted by the type checker.
This class of type-checking violations reflect an underlying logical fact: none of the component functions is capable of handling such cross-cutting values, so in the general case we do not know how to handle them in the composite function.
As a result, pattern matching cases for those constructors are left to the user to provide, and are included verbatim in the generated composite function:
\begin{lstlisting}[xleftmargin=0pt,firstline=188,lastline=192]
fn K3countNodes: K3Term → Nat := Boolean.countNodes |+ Indet.countNodes |+ Ops.countNodes
| .Neg t      => 1 + K3countNodes t
| .And t₁ t₂  => 1 + K3countNodes t₁ + K3countNodes t₂
| .Or t₁ t₂   => 1 + K3countNodes t₁ + K3countNodes t₂
| .Imp t₁ t₂  => 1 + K3countNodes t₁ + K3countNodes t₂
\end{lstlisting}

However, in the previous example, the pattern matching cases for the recursive constructors all follow the same syntactic structure of their respective cases in \code{Ops.countNodes}.
This is not necessarily true for all composed functions, because recursive constructors denote values involving feature interactions between the composed components.
When those interactions follow the same syntactic structure of the components, as in the case of \code{K3countNodes}, the composition framework attempts to copy the pattern matching alternatives from the component functions if those alternatives are not explicitly provided.
In this particular example, the following command generates the exact same function as the one using explicit pattern matching alternatives for recursive constructors:
\begin{lstlisting}[xleftmargin=0pt,firstline=188,lastline=188]
fn K3countNodes: K3Term → Nat := Boolean.countNodes |+ Indet.countNodes |+ Ops.countNodes
\end{lstlisting}


The implementation of the function composition operator \code{|+} operator follows this order when generating constructors:
(1) all the explicitly provided pattern matching cases are added first,
(2) followed by pattern matching cases for the non-recursive constructors,
(3) and finally for each of the recursive constructors, a pattern matching alternative following the same syntactic structure of that in the corresponding component function is tried. If it typechecks, it is added to the generated function, otherwise it is dropped. 
After the composite function is fully generated and elaborated, any missing pattern matching cases are reported to the user as a Lean syntax error, which is consistent with the expected behavior when writing plain Lean functions with missing patterns.

\newcommand{\bool}{\code{Boolean}}
\newcommand{\nat}{\code{Nat}}
\newcommand{\stlc}{\code{STLC}}
\newcommand{\Ty}{\code{T}}
\newcommand{\Term}{\code{Term}}
\newcommand{\Val}{\code{Val}}
\newcommand{\TRel}{\code{TRel}}
\newcommand{\countn}{\code{countNodes}}

\begin{figure}[t]
    \vspace{-0.15in}
    \begin{subfigure}[t]{0.49\textwidth}
\begin{lstlisting}[xleftmargin=0pt,firstline=4,lastline=30]
namespace Boolean
inductive T where
| Bool

inductive Term where
| True | False
| If (c t₁ t₂: Term)

@[simp] def countNodes: Term → Nat
| .True => 1
| .False => 1
| .If c t₁ t₂ => 1 + countNodes c + countNodes t₁ + countNodes t₂

inductive Val: Term → Prop
| T: Val .True | F: Val .False

inductive TRel: Term → T → Prop
| TT: TRel .True .Bool
| FF: TRel .False .Bool
| If: TRel c .Bool → TRel t₁ τ → TRel t₂ τ → TRel (.If c t₁ t₂) τ

theorem notEmpty(t: Term):countNodes t > 0 := by
  induction t with
  | True => simp
  | False => simp
  | If => simp[Nat.succ_add]
end Boolean
\end{lstlisting}
    \vspace{-0.15in}
    \centering
    \caption{\bool~type definition.}
    \label{fig:boolean}
    \end{subfigure}
    \hfill
    \begin{subfigure}[t]{0.49\textwidth}
\begin{lstlisting}[xleftmargin=0pt,firstline=4,lastline=32]
namespace Nat
inductive T where
| N

inductive Term where
| Zero
| Succ (t: Term)
| Pred (t: Term)

@[simp] def countNodes: Term → Nat
| .Zero => 1
| .Succ t => 1 + countNodes t
| .Pred t => 1 + countNodes t

inductive Val: Term → Prop
| Z: Val .Zero
| S (v: Term): Val v → Val (.Succ v)

inductive TRel: Term → T → Prop where
| Z: TRel .Zero .N
| S: TRel t .N → TRel (.Succ t) .N
| P: TRel t .N → TRel (.Pred t) .N

theorem notEmpty(t: Term): countNodes t > 0 := by
  induction t with
  | Zero => simp
  | Succ t' ih => simp[Nat.succ_add]
  | Pred t' ih => simp[Nat.succ_add]
end Nat
\end{lstlisting}
    \vspace{-0.15in}
    \centering
    \caption{\nat~type definition.}    
    \label{fig:nat}
    \end{subfigure}
    \vspace{-0.1in}
    \caption{Separate definitions of \bool~and \nat. For each of the components, the set of types \Ty, the set of syntactic terms \Term, a function \countn~for counting the number of AST nodes in a term, the set of values \Val, a type relation \TRel, and a theorem (\code{notEmpty}) are all shown.}
    \label{fig:components}
    \vspace{-0.3in}
\end{figure}

\begin{figure}[t]
\begin{lstlisting}[xleftmargin=0pt,firstline=4,lastline=38]
namespace STLC
inductive T: Type
| Fn (τ₁ τ₂: T)

abbrev Var := String

abbrev Context := Var → T
def augment (Γ: Context) (x: Var) (τ: T): Context := λv ↦ if v=x then τ else Γ v

inductive Term where
| V   (x: Var)
| Abs (x: Var) (τ: T) (b: Term)
| App (t₁ t₂: Term)

@[simp] def countNodes: Term → Nat
| .V _       => 2
| .Abs _ _ b => 3 + countNodes b
| .App t₁ t₂ => 1 + countNodes t₁ + countNodes t₂

inductive Val: Term → Prop
| A (x: Var) (τ: T) (b: Term): Val (.Abs x τ t)

inductive TRel: Context → Term → T → Prop where
| V (x: Var) (τ: T): Γ x = τ → TRel Γ (.V x) τ
| Abs (x: Var) (b: Term) (τ₁ τ₂: T):
    TRel (augment Γ x τ₁) b τ₂ → TRel Γ (.Abs x τ₁ b) (.Fn τ₁ τ₂)
| App (t₁ t₂: Term) (τ₁ τ₂: T):
    TRel Γ t₁ (.Fn τ₁ τ₂) → TRel Γ t₂ τ₁ → TRel Γ (.App t₁ t₂) τ₂

theorem notEmpty(t: Term): countNodes t > 0 := by
  induction t with
  | V => simp
  | Abs x τ b ih => apply Nat.lt_add_left; apply ih
  | App t₁ t₂ ih₁ ih₂ => simp[Nat.succ_add]
end STLC
\end{lstlisting}
    \vspace{-0.2in}
    \caption{Definition of the \stlc~component, including artifacts respective to the ones in Fig.~\ref{fig:components}.}
    \label{fig:stlc}
    \vspace{-0.2in}
\end{figure}

\vspace{-0.1in}
\section{Case Study}
\label{sec:casestudy}
\vspace{-0.1in}

The primary goal of this case study is to demonstrate how independently defined types and function definitions can be composed transparently using the type composition framework presented in this paper.
Another goal is to qualitatively evaluate the amount of manual effort needed to achieve type composition, and to identify any practical limitations of the framework.

The case study is based on the Typed Lambda Calculus (TLC) presentation from~\cite{Pierce:2002} (Chapters 8 \& 9). 
Instead of developing a single monolithic definition of TLC, our goal is to define three separate sub-languages, and then compose them.
Fig.~\ref{fig:components} and Fig.~\ref{fig:stlc} list the definitions of the three sub-languages: boolean expressions \bool~(Fig.~\ref{fig:boolean}), natural numbers \nat~(Fig.~\ref{fig:nat}), and Simply Typed Lambda Calculus \stlc~(Fig.~\ref{fig:stlc}).
Each of the sub-languages includes inductive types for the set of relevant types (\Ty), the set of valid syntactic terms (\Term), a function for counting the number of nodes in the abstract syntax tree of a term (\countn), a unary predicate (\Val) indicating whether a term is a value or not, and a type relation (\TRel) between terms and types.

The \bool~definition (Fig.~\ref{fig:boolean}) introduces a single type \code{Bool} with three syntactic categories: \code{True}, \code{False}, and the conditional term \code{If}.
Out of those, the predicate \Val~designates \code{True} and \code{False} as values.
The function \countn~returns the value 1 both for \code{True} and \code{False}, and recursively counts the nodes in the subterms of a conditional expression.
The type relation \TRel~assigns the only available type \bool~to both the \code{True} and \code{False} values. 
The type of an \code{If} term is the type of both the \emph{then} and \emph{else} terms, assuming the condition is of type \bool.

Similarly, the \nat~component (Fig.~\ref{fig:nat}) defines a single type \code{N} with three constructors: \code{Zero}, \code{Succ}, and \code{Pred}. 
Values are the term \code{Zero}, and any term composed of only the \code{Zero} and \code{Succ} constructors.
The type relation assigns the type \code{N}~to \code{Zero}, and the successors and predecessors of terms of type \code{N}.

The STLC component (Fig.~\ref{fig:stlc}) defines the plain simply-typed lambda calculus with no built-in types.
The only type introduced in this component is the \emph{arrow} type $\tau_1 \to \tau_2$ indicating a function from type $\tau_1$ to $\tau_2$.
The three syntactic terms of STLC are variables (encoded here as Lean \code{String} objects), lambda abstractions, and lambda applications.
The only category of designated values is lambda abstractions.
Because syntactic variables are introduced in this component, the type relation has an extra parameter: the variable \code{Context}, which assigns types to variables.
The variable context is used whenever a type is assigned to a syntactic category using a variable (particularly \code{V} and \code{Abs}).
In addition, \code{Abs} extends the context (using the \code{augment} function) by mapping the newly bound lambda abstraction variable to its type in the new context.

To compose the three components, we start with \Ty~(as discussed in Sec.~\ref{sec:typeComp}):  
\begin{lstlisting}[xleftmargin=0pt,firstline=6,lastline=6]
inductive T := Boolean.T |+ Nat.T |+ STLC.T
\end{lstlisting}


Composing terms is different in two respects: 
First, terms are recursive, so the recursive references in the \code{If} constructor for example are adjusted to refer to the new composite \code{Term} type, not \code{Boolean.Term}.
In addition, a new constructor \code{isZero} that crosscuts the \bool~component (its return type is \code{Bool}), and the \nat~component (its parameter is \code{N}) has to be added:
\begin{lstlisting}[xleftmargin=0pt,firstline=21,lastline=22]
inductive Term := Boolean.Term |+ Nat.Term |+ STLC.Term
| isZero (t: Term)
\end{lstlisting}



To compose the three \countn~functions, the \code{fn} is used.
Because the recursive cases follow the syntactic structure of those in the component functions, they are safely included in the composite function.
We still need to add an extra pattern matching case for \code{isZero} though.
\begin{lstlisting}[xleftmargin=0pt,firstline=72,lastline=73]
fn countNodes: Term → Nat := Boolean.countNodes |+ Nat.countNodes |+ STLC.countNodes
| Term.isZero t => 1 + countNodes t
\end{lstlisting}


Composing the \code{TRel} inductive predicates is no different from composing inductive value types in principle.
However, \code{STLC.augment} is not directly reusable, because its type is \code{STLC.Context → STLC.Var → STLC.T → STLC.Context}.
Similarly, \code{STLC.Context} is defined as \code{STLC.Var → STLC.T}. 
All references to \code{STLC.T} need to point to \code{T} instead.
This can be achieved for \code{STLC.augment} by including it in a singleton composition, which would simply adjust the names of components to the names of their corresponding composites.
Similarly, a new \code{Context} type can be defined as a supertype of \code{STLC.Context}.
\begin{lstlisting}[xleftmargin=0pt,firstline=207,lastline=212]
def Context := STLC.Var → T

instance: SubType STLC.Context Context where
  coe f := λ x ↦ f x

fn augment: Context → STLC.Var → T → Context := STLC.augment
\end{lstlisting}

Note that the coercion operator from \code{STLC.Context} to \code{Context} (in the \code{SubType} instance) transparently invokes the coercion operator from \code{STLC.T} to \code{T} when returning the result of the context function.

Now the individual type relations can be composed into one, adding a new constructor for the \code{isZero} syntactic category:
\begin{lstlisting}
inductive TRel: Context → Term → T → Prop := Boolean.TRel |+ Nat.TRel |+ STLC.TRel
| iz: TRel Γ t T.N → TRel Γ (Term.isZero t) T.Bool
\end{lstlisting}

Here the type is explicitly specified as \code{Context → Term → T → Prop} 
because \code{Boolean.TRel} and \code{Nat.TRel} do not take a \code{Context} argument.
The result when printing the generated \code{TRel} is:
\begin{lstlisting}
inductive TRel : Context → Term → T → Prop
number of parameters: 0
constructors:
TRel.TT : ∀ {_ : Context}, TRel _ Term.True T.Bool
TRel.FF : ∀ {_ : Context}, TRel _ Term.False T.Bool
TRel.If : ∀ {_ : Context} {c t1 : Term} {τ : T} {t2 : Term},
TRel _ c T.Bool → TRel _ t1 τ → TRel _ t2 τ → TRel _ (c.If t1 t2) τ
TRel.Z : ∀ {_ : Context}, TRel _ Term.Zero T.N
TRel.S : ∀ {_ : Context} {t : Term}, TRel _ t T.N → TRel _ t.Succ T.N
TRel.P : ∀ {_ : Context} {t : Term}, TRel _ t T.N → TRel _ t.Pred T.N
TRel.V : ∀ Γ{ : Context} (x : STLC.Var) (τ : T), Γ x = τ → TRel Γ (Term.V x) τ
TRel.Abs : ∀ Γ{ : Context} (x : STLC.Var) (b : Term) (τ 1 τ 2 : T),
TRel (augment Γ x τ 1) b τ 2 → TRel Γ (Term.Abs x τ 1 b) (τ 1.Fn τ 2)
TRel.App : ∀ Γ{ : Context} (t1 t2 : Term) (τ 1 τ 2 : T), TRel Γ t1 (τ 1.Fn τ 2) → TRel Γ t2 τ 1 → TRel Γ (t1.App t2) τ 2
TRel.iz : ∀ Γ{ : Context} {t : Term}, TRel Γ t T.N → TRel Γ t.isZero T.Bool    
\end{lstlisting}

Implicit parameters of type \code{Context} were added to constructors that didn't originally take a \code{Context} argument (e.g., \code{TRel.TT}).
Similarly, holes (\code{\_}) were added to the predicate terms in place of the added \code{Context} parameter.
\code{Context} and \code{augment} are also used instead of \code{STLC.Context} and \code{STLC.augment}.

In Lean, and the Calculus of Inductive Constructions (CIC) in general, a theorem is a type, and its proof is a term of that type.
This is a direct result of the Curry-Howard isomorphism applied to constructive logics.
As a result, a theorem with a set of premises and a conclusion is a function type mapping proofs of the premises to a proof of the conclusion.
A proof is a function of that type.
If proofs of theorems defined on component types already exist, they can be composed exactly the same way we compose functions.
For example, to compose the \code{notEmpty} component theorems into a composite theorem, the same notation for function composition is used:
\begin{lstlisting}[xleftmargin=0pt,firstline=104,lastline=110]
fn notEmpty (t: Term) : countNodes t > 0 := Boolean.notEmpty |+ Nat.notEmpty |+ STLC.notEmpty
| .If c t e   => by simp[countNodes, Nat.succ_add]
| .Succ t     => by simp[countNodes, Nat.succ_add]
| .Pred t     => by simp[countNodes, Nat.succ_add]
| .Abs v τ t  => by simp[countNodes]; apply Nat.lt_add_left; apply notEmpty t
| .App t₁ t₂  => by simp[countNodes, Nat.succ_add]
| .isZero t   => by simp[countNodes, Nat.succ_add]
\end{lstlisting}

In this particular example the structure of the component proof objects is not a pattern matching on terms, but instead a more complex Lean term involving the recursor on term objects.
As a result, the pattern matching cases explicitly included for recursive constructors cannot be copied from component proofs.

\vspace{0.1in}
\heading{Limitations.}
While working on this case study, some limitations of the composition framework were identified:
(1) Not all Lean types are supported. For example, nested inductive types and inductive families are not currently supported.
Polymorphic and dependent types are supported though.
(2) Composing mutually recursive functions is not currently supported. 
Adding support for mutual recursion can be added with some engineering effort, possibly by building a function-call graph.
Functions forming a cycle in the call graph are mutually recursive, and should be rewritten together within a single \code{mutual} block.
(3) In this case study each of the components was implemented within a separate \code{namespace}, and type and function definitions that correspond to the artifacts to be composed were consistently named.
However, the composition framework does not currently leverage this, and we had to issue separate commands for composing \Ty, \Term, \countn, \Val, and \TRel. Ideally, it would make more sense to compose \emph{modules} and all the definitions they encompass, instead of composing individual definitions.
In general, whether definitions are grouped within a module or a namespace, the framework can be extended to match composable artifacts, either using their names or type signatures, and composing them collectively in one shot.
(4) Sometimes non-inductive types (such as the \code{STLC.Context} type) need to be modified. In the case study a solution was hacked by adding an explicit instance of \code{SubType}. 
In the future this can be automated by traversing a dependency graph and automatically rewriting definitions accordingly.
(5) As demonstrated by the composition of the \code{notEmpty} proofs, symbolic generation of the recursive cases is not currently supported.
For future work, we are considering several approaches to supporting composite proofs.
Extending the framework along the lines of~\cite{Boite:2004} by attempting to refine a composition of proof objects is one possible direction to pursue.
Another is analyzing the proof scripts used to prove theorems of the individual components and identifying tactic usage patterns that can be reused when proving composite theorems.
\vspace{-0.15in}
\section{Related Work}
\label{sec:related}
\vspace{-0.15in}

Boite~\cite{Boite:2004} presented a set of Coq syntax extensions and tactics for extending inductive types with extra constructors and type parameters. 
His constructs also syntactically reuse as much as possible of function definitions and theorem proofs depending on the original type definition.
When composing proofs, a special tactic is used to manipulate the component proof objects, and generate a derived proof for the composite theorem.
This approach does not utilize subtyping at all, which is the main difference between it and our approach.
In addition, it only extends a given type with extra constructors, and does not combine multiple existing inductive types.

Modular composition of definitions, theorems and proofs into feature-based product lines was presented in~\cite{Delaware:2011}.
This approach composes feature modules explicitly parameterized by \term{Variation Points (VPs)}.
As a result, a component has to anticipate how other components might interact with it, and syntactically provide those VP parameters.
Our approach on the other hand does not require components to anticipate interactions, so VPs are not needed, and we can compose plain Lean definitions.
In addition, modules for feature interactions of different arities have to be composed together with interacting components.
Although we require feature interactions to be explicitly defined, they are defined only once at composition time for the entire set of composed components, instead of separate pairwise and higher arity interactions.

The data types a-la-carte (DTAC) work for Haskell~\cite{Swierstra:2008} composes functions defined on individual constructor signatures, provided those signatures are implemented as functors.
Composition relies on the fact that coproducts of functors are functors.
Again, this approach requires components to be implemented in a particular way, and does not support composing plain inductive types.

Because of the termination guarantees of proof languages based on the Calculus of Inductive Constructions (e.g., Coq, Lean), the fixpoint definitions of DTAC are not allowed.
Meta-theory a la carte~\cite{Delaware:2013} adapts DTAC to Coq by using Church encodings of types, and Mendler-style folds on them.
Since our approach relies on copying constructor definitions from components to composites, and both upward and downward coercions, we do not use functors and folds.

Coq-a-la-carte~\cite{Forster:2020} uses metaprogramming (MetaCoq and Autosubst 2) to automatically generate much of the boilerplate code for functors, functor coproducts, function algebras and typeclass instances.
Stronger induction principles of composite types are also automatically generated.
Still, component types are expected to be written as Higher-Order Abstract Syntax (HOAS) definitions, and component-level functions are written over functors.
This is the primary difference between their approach and ours.
As we focus on reusing plain Lean definitions, we do not assume those components to be aware of the composition mechanism, such as using HOAS definitions or component functors.
Our approach is more simplistic, but this simplicity alleviates much of the challenges addressed by their approach, such as stronger induction principles and inductive predicates (we get both almost for free).

Jin et al.\cite{Jin:2023} developed a Coq language extension (FPOP) with family polymorphism as a Coq plugin. They extend the Martin-Löf dependent type theory with family polymorphism, and demonstrate their solution on some case studies for modular composition of formal language modules. Their approach relies on Object-Oriented style inheritance and polymorphic late binding, while our approach is based on subtyping.
In addition, components have to be written in FPOP, while we support composing plain Lean inductive types, functions, and theorems. 




\vspace{-0.15in}
\section{Conclusion}
\label{sec:conclusion}
\vspace{-0.15in}

Extensibility and reuse of inductive data types (which are closed by design) is a challenge in many programming and theorem proving languages.
In this paper we presented a set of algorithms for composing inductive data types, inductive predicates, and implementations of pattern matching functions.
Our approach uses Lean meta-programming to analyze the syntactic structure of existing definitions and syntactically compose them into new ones.

We presented high level algorithms for inductive data type and function implementation composition.
We also discussed the implementation of those algorithms in Lean.
In addition, subtyping relations between constituent and composite types were discussed both at the type theoretic and implementation levels.
A case study was discussed, highlighting both the features and limitations of the composition framework.

For future work, we plan to investigate approaches to improve proof composition.
We also plan to study intuitive syntactic notations that would allow partial reuse of sub-proofs while allowing users to augment them with proofs of additional subgoals.

\begin{credits}
\subsubsection{\ackname}
The author would like to thank anonymous reviewers for their insightful feedback and suggestions.

\subsubsection{\discintname}
The author has no competing interests to declare that are relevant to the content of this paper.
\end{credits}
\bibliographystyle{splncs04}
\bibliography{pl,se,spl,logic}

@Inbook{Hazen:2019,
author="Hazen, Allen P.
and Pelletier, Francis Jeffry",
title="K3, {\L}3, LP, RM3, A3, FDE, M: How to Make Many-Valued Logics Work for You",
bookTitle="New Essays on Belnap-­Dunn Logic",
year="2019",
publisher="Springer International Publishing",
address="Cham",
pages="155--190",
isbn="978-3-030-31136-0",
doi="10.1007/978-3-030-31136-0_11",
url="https://doi.org/10.1007/978-3-030-31136-0_11"
}

@InProceedings{Pfenning:1990,
author="Pfenning, Frank
and Paulin-Mohring, Christine",
editor="Main, M.
and Melton, A.
and Mislove, M.
and Schmidt, D.",
title="Inductively defined types in the Calculus of Constructions",
booktitle="Mathematical Foundations of Programming Semantics",
year="1990",
publisher="Springer-Verlag",
address="New York, NY",
pages="209--228",
isbn="978-0-387-34808-7"
}

@article{Delaware:2013,
author = {Delaware, Benjamin and d. S. Oliveira, Bruno C. and Schrijvers, Tom},
title = {Meta-theory \`{a} la carte},
year = {2013},
issue_date = {January 2013},
publisher = {Association for Computing Machinery},
address = {New York, NY, USA},
volume = {48},
number = {1},
issn = {0362-1340},
url = {https://doi.org/10.1145/2480359.2429094},
doi = {10.1145/2480359.2429094},
journal = {SIGPLAN Not.},
month = jan,
pages = {207–218},
numpages = {12}
}

@inproceedings{deMoura:2021,
	address={Cham}, 
	title={{The Lean 4 Theorem Prover and Programming Language}}, 
	ISBN={978-3-030-79876-5}, 
	abstractNote={Lean 4 is a reimplementation of the Lean interactive theorem prover (ITP) in Lean itself. It addresses many shortcomings of the previous versions and contains many new features. Lean 4 is fully extensible: users can modify and extend the parser, elaborator, tactics, decision procedures, pretty printer, and code generator. The new system has a hygienic macro system custom-built for ITPs. It contains a new typeclass resolution procedure based on tabled resolution, addressing significant performance problems reported by the growing user base. Lean 4 is also an efficient functional programming language based on a novel programming paradigm called functional but in-place. Efficient code generation is crucial for Lean users because many write custom proof automation procedures in Lean itself.},
	booktitle={Automated Deduction – CADE 28}, 
	publisher={Springer International Publishing}, 
	author={Moura, Leonardo de and Ullrich, Sebastian}, 
	editor={Platzer, André and Sutcliffe, Geoff}, 
	year={2021}, 
	pages={625–635} }

@inproceedings{Forster:2020,
author = {Forster, Yannick and Stark, Kathrin},
title = {Coq \`{a} la carte: a practical approach to modular syntax with binders},
year = {2020},
isbn = {9781450370974},
publisher = {Association for Computing Machinery},
address = {New York, NY, USA},
url = {https://doi.org/10.1145/3372885.3373817},
doi = {10.1145/3372885.3373817},
booktitle = {Proceedings of the 9th ACM SIGPLAN International Conference on Certified Programs and Proofs},
pages = {186–200},
numpages = {15},
location = {New Orleans, LA, USA},
series = {CPP 2020}
}

@article{Jin:2023,
author = {Jin, Ende and Amin, Nada and Zhang, Yizhou},
title = {Extensible Metatheory Mechanization via Family Polymorphism},
year = {2023},
issue_date = {June 2023},
publisher = {Association for Computing Machinery},
address = {New York, NY, USA},
volume = {7},
number = {PLDI},
url = {https://doi.org/10.1145/3591286},
doi = {10.1145/3591286},
journal = {Proc. ACM Program. Lang.},
month = jun,
articleno = {172},
numpages = {25}
}

@inproceedings{Klein:2009,
    title={{seL4: Formal verification of an OS kernel}},
    author={Klein, Gerwin and Elphinstone, Kevin and Heiser, Gernot and Andronick, June and Cock, David and Derrin, Philip and Elkaduwe, Dhammika and Engelhardt, Kai and Kolanski, Rafal and Norrish, Michael and others},
    booktitle={Proceedings of the ACM SIGOPS 22nd Symposium on Operating Systems Principles},
    pages={207--220},
    year={2009}
}

@article{Leroy:2009,
    author = {Xavier Leroy},
    title = {{A Formally Verified Compiler Back-end}}, 
    journal = {Journal of Automated Reasoning},
    volume = 43,
    number = 4,
    pages = {363--446},
    year = 2009,
    url = {http://xavierleroy.org/publi/compcert-backend.pdf},
    urlpublisher = {http://dx.doi.org/10.1007/s10817-009-9155-4},
    hal = {http://hal.inria.fr/inria-00360768/},
    pubkind = {journal-int-mono}
}

@book{Pierce:2002,
	author = {Pierce, Benjamin C.},
	title = {{Types and Programming Languages}},
	year = {2002},
	isbn = {0262162091, 9780262162098},
	edition = {1st},
	publisher = {The MIT Press},
}

@article{Swierstra:2008,
author = {Swierstra, Wouter},
title = {Data types \`{a} la carte},
year = {2008},
issue_date = {July 2008},
publisher = {Cambridge University Press},
address = {USA},
volume = {18},
number = {4},
issn = {0956-7968},
url = {https://doi.org/10.1017/S0956796808006758},
doi = {10.1017/S0956796808006758},
journal = {J. Funct. Program.},
month = jul,
pages = {423–436},
numpages = {14}
}

@article{Wadler:1998,
  title={{The Expression Problem}},
  author={Wadler, Philip},
  journal={Posted on the Java Genericity mailing list},
  year={1998},
  url = {https://homepages.inf.ed.ac.uk/wadler/papers/expression/expression.txt}
}

@inproceedings{Batory:2011,
author = {Batory, Don and H\"{o}fner, Peter and Kim, Jongwook},
title = {Feature interactions, products, and composition},
year = {2011},
isbn = {9781450306898},
publisher = {Association for Computing Machinery},
address = {New York, NY, USA},
url = {https://doi.org/10.1145/2047862.2047867},
doi = {10.1145/2047862.2047867},
booktitle = {Proceedings of the 10th ACM International Conference on Generative Programming and Component Engineering},
pages = {13–22},
numpages = {10},
location = {Portland, Oregon, USA},
series = {GPCE '11}
}

@InProceedings{Boite:2004,
author="Boite, Olivier",
editor="Slind, Konrad
and Bunker, Annette
and Gopalakrishnan, Ganesh",
title={{Proof Reuse with Extended Inductive Types}},
booktitle="Theorem Proving in Higher Order Logics",
year="2004",
publisher="Springer Berlin Heidelberg",
address="Berlin, Heidelberg",
pages="50--65",
isbn="978-3-540-30142-4"
}

@Inbook{Dijkstra:1982,
author={{Dijkstra, Edsger W.}},
title="On the Role of Scientific Thought",
bookTitle="Selected Writings on Computing: A personal Perspective",
year="1982",
publisher="Springer New York",
address="New York, NY",
pages="60--66",
isbn="978-1-4612-5695-3",
doi="10.1007/978-1-4612-5695-3_12",
url="https://doi.org/10.1007/978-1-4612-5695-3_12"
}

@article{Harrison:1993,
author = {Harrison, William and Ossher, Harold},
title = {Subject-oriented programming: a critique of pure objects},
year = {1993},
issue_date = {Oct. 1, 1993},
publisher = {Association for Computing Machinery},
address = {New York, NY, USA},
volume = {28},
number = {10},
issn = {0362-1340},
url = {https://doi.org/10.1145/167962.165932},
doi = {10.1145/167962.165932},
journal = {SIGPLAN Not.},
month = oct,
pages = {411–428},
numpages = {18}
}

@InProceedings{Kiczales:1997,
author="Kiczales, Gregor
and Lamping, John
and Mendhekar, Anurag
and Maeda, Chris
and Lopes, Cristina
and Loingtier, Jean-Marc
and Irwin, John",
title="Aspect-oriented programming",
booktitle="ECOOP'97 --- Object-Oriented Programming",
year="1997",
publisher="Springer Berlin Heidelberg",
address="Berlin, Heidelberg",
pages="220--242",
isbn="978-3-540-69127-3"
}

@article{Parnas:1972,
author = {{Parnas, D. L.}},
title = {On the criteria to be used in decomposing systems into modules},
year = {1972},
issue_date = {Dec. 1972},
publisher = {Association for Computing Machinery},
address = {New York, NY, USA},
volume = {15},
number = {12},
issn = {0001-0782},
url = {https://doi.org/10.1145/361598.361623},
doi = {10.1145/361598.361623},
journal = {Commun. ACM},
month = dec,
pages = {1053–1058},
numpages = {6}
}

@inproceedings{Delaware:2011,
    author = {Delaware, Benjamin and Cook, William and Batory, Don},
    title = {{Product Lines of Theorems}},
    booktitle = {Proceedings of the 2011 ACM International Conference on Object Oriented Programming Systems Languages and Applications},
    series = {OOPSLA '11},
    year = {2011},
    isbn = {978-1-4503-0940-0},
    location = {Portland, Oregon, USA},
    pages = {595--608},
    numpages = {14},
    url = {http://doi.acm.org/10.1145/2048066.2048113},
    doi = {10.1145/2048066.2048113},
    acmid = {2048113},
    publisher = {ACM},
    address = {New York, NY, USA},
}

\end{document}